\documentclass{article}
\usepackage{axodraw}
\input{epsf}
\newcommand{\beq}{\begin{equation}}
\newcommand{\eeq}{\end{equation}}
\newcommand{\beqa}{\begin{eqnarray}}
\newcommand{\eeqa}{\end{eqnarray}}

\newcommand{\tdz}{\tilde{\delta}_0}
\newcommand{\tdt}{\tilde{\delta}_2}
\newcommand{\tdg}{\tilde{\delta}_\gamma}
\newcommand{\bdz}{\bar{\delta}_0}
\newcommand{\bdt}{\bar{\delta}_2}
\newcommand{\bdg}{\bar{\delta}_\gamma}
\newcommand{\Dz}{{\tilde{\delta}_0-\bar{\delta}_0}}
\newcommand{\Dt}{{\tilde{\delta}_2-\bar{\delta}_2}}
\newcommand{\Dg}{{\tilde{\delta}_\gamma-\bar{\delta}_\gamma}}
\newcommand{\Dez}{\Delta_0}
\newcommand{\Det}{\Delta_2}
\newcommand{\Deg}{\Delta_{\gamma}}

\newcommand{\vs}{\vspace{-0.20cm}}
\begin{document}


\begin{flushright}
{\tiny  FZJ-IKP(TH)-2001-05} \\
{\tiny  UK/TP-2001-02} \\
\end{flushright}

\vspace{1in}

\begin{center}

\bigskip

{{\Large\bf Watson's theorem and electromagnetism in 
    \boldmath{$ K \to \pi\pi$} decay
    }}

\end{center}

\vspace{.3in}

\begin{center}
{\large 
S. Gardner$^\dagger$\footnote{email: gardner@pa.uky.edu},
Ulf-G. Mei{\ss}ner$^\ddagger$\footnote{email: u.meissner@fz-juelich.de},
G. Valencia$^\star$\footnote{email: valencia@iastate.edu}}

\vspace{1cm}

$^\dagger${\it Department of Physics and Astronomy, University of Kentucky\\
Lexington, Kentucky 40506-0055, USA}\\

\bigskip

$^\ddagger${\it Forschungszentrum J\"ulich, Institut f\"ur Kernphysik 
(Theorie)\\ D-52425 J\"ulich, Germany}

\bigskip

$^\star${\it Department of Physics and Astronomy, Iowa State
  University\\ Ames, Iowa 50011, USA}

\bigskip

\end{center}

\vspace{.6in}

\thispagestyle{empty} 

\begin{abstract}\noindent 
We consider what constraints unitarity and CPT invariance 
yield on the strong and electromagnetic phases entering
$K\to \pi\pi$ decay. 
In particular, we show that the relative size of the 
electromagnetically-induced changes in the 
$I=0$ and $I=2$ phase shifts 
in the two--pion final state do not
depend on the explicit coupling to the $\pi^+\pi^-\gamma$ channel.  
This demonstrates that Watson's theorem can be extended
to include the presence of electromagnetism. 
We point out the consequences for the 
general structure of the $K\to\pi\pi$ 
decay amplitudes in the presence of
isospin violation.
\end{abstract}

\vfill

\pagebreak

\noindent {\bf 1.} 
A detailed understanding of the rich phenomenology of 
$K \to \pi\pi$ decays has remained elusive despite decades of
effort. Although progress has been made, 
the dynamical origin of the $\Delta I =1/2$ rule, as well as 
the strength of CP-violating parameter Re~$(\epsilon'/\epsilon)$, 
is not yet clear. Another,
presumably related, puzzle stems from the apparent violation of 
Watson's final-state theorem. 
Watson's theorem emerges from unitarity and CPT-invariance,
in concert with isospin symmetry, and implies that
the strong phase in $K\to \pi\pi$ decay ought be given by 
that of $\pi\pi$ scattering. However,
the S--wave $\pi\pi$ phase shift difference 
$\delta_0 - \delta_2$ extracted from the $K\to \pi\pi$ decay 
modes, using physical masses in the phase-space integrals, is about 
$57^\circ$~\cite{GMI}, whereas its value from 
$\pi\pi$ scattering data, with the help of 
chiral perturbation theory and dispersion relations, is about 
%
45$^\circ$ with an uncertainty of roughly 10\%~\cite{GMI,Bern}. 
The assumed equality of these quantities is a consequence of 
isospin symmetry, so that the resolution of the discrepancy 
has been sought in the computation of isospin-violating effects.
Isospin violation can be generated by both strong (up-down 
quark mass difference) 
and electromagnetic 
(virtual photon) interactions, 
and its effects have been recently studied in 
great detail, see, e.g., 
Refs.\cite{EdR}-\cite{WienII}. 
While many interesting 
results have been obtained 
and many others are forthcoming,
the gap between the phase-shift difference obtained from 
$K\to \pi\pi$ decay and
$\pi\pi$ scattering has thus far eluded a detailed explanation. 
In the
framework of chiral perturbation theory, which is the appropriate theoretical
tool in this context, 
many new low--energy constants appear 
in the most general Lagrangian of Goldstone bosons coupled to 
virtual photons and external sources, making certain numerical predictions
difficult. 
It is thus important to explore 
whether Watson's theorem can be extended in the presence of isospin
violation. In Ref.~\cite{GV} Watson's theorem was shown to persist
through leading order in the up-down quark mass difference, so that 
the phase shifts from 
$K\to \pi\pi$ decay and
$\pi\pi$ scattering ought be equal to ${\cal O}((m_d-m_u)^2)$. 
Were the phase shifts from 
$K\to \pi\pi$ decay and
$\pi\pi$ scattering equal, the empirical phase-shift discrepancy
could nevertheless be resolved, for it 
could be interpreted in terms of an 
additional amplitude in $K\to\pi\pi$ decay, of
$|\Delta I|=5/2$ in character~\cite{GV}. In the isospin-perfect limit,
the $K\to\pi\pi$ transition in ${\cal O}(G_F)$ can be of 
$|\Delta I|=1/2$ or $|\Delta I|=3/2$. In the presence
of isospin violation, a 
$|\Delta I|=5/2$ transition can be realized from 
$m_d \ne m_u$ effects in concert with a 
$|\Delta I|=3/2$ weak transition or from electromagnetic 
effects in concert with a 
$|\Delta I|=1/2$ weak transition. 
The empirical enhancement
of the $|\Delta I|=1/2$ weak transition suggests that
the latter mechanism is of greater importance. 
The empirical $|\Delta I|=5/2$ amplitude required to resolve
the phase-shift discrepancy is compatible with that expected
from electromagnetic effects~\cite{GV}, yet it is significantly larger
in magnitude than and of opposite sign to that indicated by
explicit estimates~\cite{CDGI,CDGII}. Moreover, including 
the estimated 
phase-shift difference from electromagnetism~\cite{CDGI,CDGII} 
exacerbates this discrepancy. 
These difficulties prompt the consideration of 
Watson's theorem in the presence of
electromagnetism, in order to realize what  
constraints may exist 
on the strong and electromagnetically induced phase shifts
in $K\to\pi\pi$ decay. 
This is the aim of the present investigation. It has been triggered by 
the work
of Bernstein~\cite{aron}, who considered isospin violation in near-threshold
neutral pion photoproduction from protons, extending the final-state theorem
to the situation of three open channels, in that case $\gamma p$, $\pi^0 p$, 
and $\pi^+ n$, where $p \,(n)$ denotes the proton (neutron). 
In a similar fashion,
we consider three open channels for the $K^0$ decays, which are the two--pion
final states with total isospin zero and two, 
denoted as $(\pi\pi)_0$ and $(\pi\pi)_2$, respectively, 
as well as the inelastic $\pi^+\pi^-\gamma$ channel, whose
inclusion is required to render the electromagnetic
corrections to the $K\to (\pi\pi)_I$ amplitudes
infrared (IR) finite. We will construct
a general 4$\times$4 S--matrix appropriate to this scenario
and derive a set of unitarity constraints from it. Our purpose 
is not a detailed 
numerical analysis of the various isospin--violating effects, 
but rather the construction of a 
theoretical framework which would be helpful in 
constraining such calculations. 
Nevertheless,
we will be able to derive consequences 
from the unitarity constraints which
thus far have only appeared indirectly in 
numerical analyses.

\medskip

\noindent {\bf 2.} 
First, we must collect some definitions for
the discussion of the $K \to \pi\pi$ amplitudes  --- we follow 
the notation and conventions of 
Ref.~\cite{GV} and refer the reader to that paper for further details. 
In the isospin limit ($m_u = m_d, e=0$),
the decay of a neutral kaon into two pions with isospin $I$ equal
to  zero
or two can be parametrized via\footnote{Here $A_I$ is $i$ times
$A_I$ defined in the Ref.~\cite{GV}.}
\beqa
\langle (\pi\pi)_I \, | \, {\cal H}_{\rm W} \, | \, K^0  \rangle
&=& A_I \, \exp(i \, \delta_I)~, \nonumber \\
\langle (\pi\pi)_I \, | \, {\cal H}_{\rm W} \, | \, \overline{K^0}  \rangle
&=& A_I^* \, \exp(i \, \delta_I)~, 
\eeqa
where ${\cal H}_{\rm W}$ is the effective weak Hamiltonian for kaon
decay. 
The amplitude $A_I$ is such that $A_I = |A_I| \exp(i\xi_I)$,
with $\xi_I$ the weak phase associated with the decay to the final
two--pion state of isospin $I$, and $\delta_I$ is the phase shift
corresponding to S--wave $\pi\pi$ scattering of isospin $I$. In the
isospin-symmetric limit, 
Bose symmetry requires the pion pair to have $I=0$ or
$I=2$. In that limit, the S--matrix for strong scattering in the
$(\pi\pi)_I$ final state is described by a pure phase,
\beqa
{\cal S} = \left( \matrix { e^{2i \delta_0} & 0 \nonumber \\
                            0 & e^{2i \delta_2} }  
                           \!\!\!\!\!\!\!\!\!\!\!\!\!\!\!\!\!\!\! \right) 
 \,\, .
\eeqa
Here, we have tacitly assumed 
that at $\sqrt{s} = M_{K^0}$ the inelasticities
from the opening of the four--pion threshold, $2\pi \to 4\pi$, are
negligible. This is
not only a well--known empirical fact~\cite{MMS,JLP}, but 
it can also be understood in
the framework of chiral perturbation theory, for it first occurs in
three--loop order (see Ref.~\cite{GMII}, e.g.). 

\medskip\noindent
We now turn to the  
inclusion of isospin-breaking effects. For the moment we neglect
the presence of the channel $K\to \pi^+ \pi^- \gamma$, and 
we consider merely how isospin violation impacts the 
$\pi\pi$ subspace. 
As previously, we introduce the
channels $(\pi\pi)_0$ and $(\pi\pi)_2$, where these states
are related to the physical basis 
via\footnote{Note that the properly symmetrized state 
$| \pi^-\pi^+ \rangle_{\rm sym} \equiv 
(| \pi_1^+\pi_2^- \rangle +  | \pi_1^-\pi_2^+ \rangle)/\sqrt{2}
= \sqrt{2} |\pi^+\pi^-\rangle$.}
\begin{eqnarray}
| \pi^+\pi^-\rangle &\propto& 
|(\pi\pi)_0 \rangle 
+ \frac{1}{\sqrt{2}}  |(\pi\pi)_2 \rangle 
\nonumber \\
| \pi^0\pi^0\rangle 
&\propto& 
|(\pi\pi)_0 \rangle 
- {\sqrt{2}}  |(\pi\pi)_2 \rangle 
\;.
\label{isodecomp}
\end{eqnarray}
Enforcing unitarity and time--reversal invariance,
the general S--matrix appropriate to scattering in the 
two--pion subspace --- with zero net charge ---
contains exactly three
parameters. Two parameters characterize $\pi\pi$ scattering 
in the isospin-perfect limit, so that 
the additional parameter must be at least of ${\cal
  O}(m_d-m_u)$ or ${\cal O}(e)$,
as isospin-breaking is generated by 
both strong and electromagnetic effects. 
Our neglect of the $\pi^+\pi^-\gamma$ channel would be 
appropriate were we to consider strong-interaction
isospin-violating effects only. Let us do this and
examine the extensions necessary for the treatment
of electromagnetism later. 
Working in analogy to the ``bar phase shifts'' 
for 
$J=S=1$ nucleon--nucleon (NN) scattering in the presence of a tensor 
force~\cite{preston}, we parametrize the S--matrix as 
\beqa\label{Sbar}  
{\cal S} = \left( \matrix { e^{i \bdz} & 0 \nonumber \\
                            0 & e^{i \bdt} }  
                           \!\!\!\!\!\!\!\!\!\!\!\!\!\!\!\!\!\!\!
                         \right) 
 \left( \matrix { \cos \, 2\kappa & i \, \sin\, 2\kappa \nonumber \\
                  i \, \sin\, 2\kappa &  \cos \, 2\kappa }
                           \!\!\!\!\!\!\!\!\!\!\!\!\!\!\!\!\!\!\!
                         \right) 
 \left( \matrix { e^{i \bdz} & 0 \nonumber \\
                            0 & e^{i \bdt} }  
                           \!\!\!\!\!\!\!\!\!\!\!\!\!\!\!\!\!\!\!
                         \right) ~~,
\eeqa
where $\kappa$ is the third parameter which is sensitive to 
isospin-violating effects. 
In the
absence of isospin violation, $\kappa = 0$, and we have
$\bar{\delta}_I = \delta_I$. 
G--parity arguments show that 
 strong isospin--violating effects in
$\pi\pi$--scattering are of 
${\cal O}((m_d-m_u)^2)$~\cite{GL,GV}. 

\medskip\noindent
We parametrize the $K\to
\pi\pi$ amplitudes in the presence of isospin violation via
\beqa
\langle (\pi\pi)_I \, | \, S \, | \, K^0  \rangle
&=& i A_I \, \exp(i \, \tilde{\delta}_I)~, \nonumber\\
\langle (\pi\pi)_I \, | \, S \, | \, \overline{K^0}  \rangle
&=& i A_I^* \, \exp(i \, \tilde{\delta}_I)~,
\eeqa
noting that 
the $\tilde{\delta}_I$ are the strong phases of
the $K\to \pi\pi$ amplitude and
recalling that $S = 1 +
i\,T$. Unitarity constrains the explicit 
relation between 
the $\tilde{\delta}_I$ and the
$\bar{\delta}_I$, note Ref.~\cite{GV}. 
If the channel-coupling
parameter were zero, then $\tilde{\delta}_I =\bar{\delta}_I =
\delta_I$, and 
the strong phase in the $K \to \pi\pi$
decay would be that of $\pi\pi$ scattering in the 
isospin-perfect limit. 
For later use, we introduce the abbreviation 
\beq\label{defD}
\Delta_I \equiv \tilde{\delta}_I - \bar{\delta}_I~, \quad I = 0, 2~,
\label{Deldef}
\eeq
so that $\Delta_I =0$ for $\kappa = 0$.

\medskip\noindent  
To end this discussion, we briefly return to the isospin-perfect limit.
The 3$\times$3 S--matrix describing the coupling of the $K_0$ to
the $(\pi\pi)_I$ channels takes the form, where
$K_0, (\pi\pi)_0$, and $(\pi\pi)_2$ refer to rows/columns 1,2 and 3, 
in order:
\beqa\label{master3}
{\cal S} = 
\left( \matrix { 1 & iA_0^* \, e^{i\tdz} &  iA_2^* \, e^{i\tdt} 
     \nonumber \\ 
                 iA_0 \, e^{i\tdz} &  e^{2i\delta_0} & 0
     \nonumber \\
                 iA_2 \, e^{i\tdt} &  0 & e^{2i\delta_2}
     \nonumber\\} \!\!\!\!\!\!\!\!\!\!\!\!\!\!\!\!\!\!  \right) 
 \,\, , 
\eeqa
so that 
${\cal S}_{21}=\langle (\pi\pi)_0 | S | K^0 \rangle$.
The amplitude $A_I$ contains a non--trivial weak phase. 
By CPT invariance, 
$\langle (\pi\pi)_I | T^\dagger | K^0 \rangle
= (\langle (\pi\pi)_I | T | \bar{ K^0} \rangle )^*$, so that
${\cal S}_{12}=\langle K^0 | S | (\pi\pi)_0 \rangle
=\langle (\pi\pi)_0 | S | \bar{ K^0} \rangle$.
We work in ${\cal O}(G_F,e^0)$, though we neglect terms
of ${\cal O}(G_F)$ in our parametrization of the
$(\pi\pi)_I \leftrightarrow (\pi\pi)_{I'}$ S--matrix.
We do this as our interest is in the 
constraints which exist on the 
T-conserving 
phases associated with an amplitude $A_I$, 
so that only the unitarity constraints emerging from 
$({\cal S}^\dagger{\cal S})_{12}=({\cal S}^\dagger{\cal S})_{13}=0$
are of interest. An amplitude $A_I$ is itself of ${\cal O}(G_F)$, so
that ${\cal O}(G_F)$ contributions to 
$(\pi\pi)_I \leftrightarrow (\pi\pi)_{I'}$ scattering play no
role in the order of $G_F$ to which we work. Thus it is appropriate
to neglect ${\cal O}(G_F)$ effects in our description of
$(\pi\pi)_I \leftrightarrow (\pi\pi)_{I'}$ scattering; we can
parametrize this $2\times 2$ matrix by a form which is 
both unitary and T-conserving. 
The unitarity constraints $({\cal S}^\dagger {\cal S})_{12} = 
({\cal S}^\dagger {\cal S})_{13} = 0$
lead to $\tilde{\delta}_I = \delta_I \,\, (I=0,2)$, so that the 
strong phases
appearing in 
$K\to \pi\pi$ decay are exactly those 
of elastic $\pi\pi$ scattering.  
This is {\em Watson's theorem} 
in the isospin--perfect world. Equipped with these
results, we are now in position to generalize this framework
to include the $\pi^+\pi^- \gamma$ final state as well.

\medskip

\noindent {\bf 3.} 
We wish to extend Watson's final--state theorem to
include both electromagnetic and strong--interaction 
isospin--violating effects in $K\to \pi\pi$ decays. For that, we extend
the  3$\times$3 matrix of Eq.~(\ref{master3}) to an appropriate matrix
of larger dimension. Before presenting this extension, let us 
collect and discuss the assumptions of our analysis. 
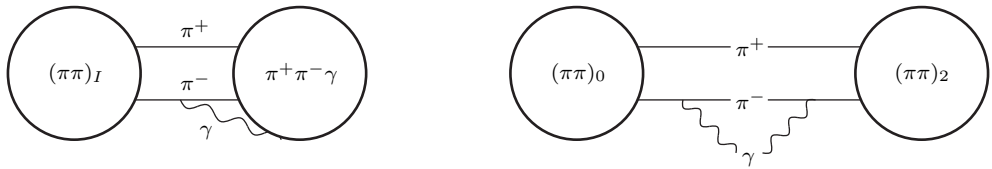
\begin{figure}[ht]         
   \hspace*{\fill} 
\begin{picture}(80,50)(-40,-15)    
\Line(-20,20)(28,20) \Text(0,27)[lc]{\footnotesize$\pi^+\;$}
\Line(-20,0)(28,0) \Text(0,7)[lc]{\footnotesize$\pi^-\;$} 
\Photon(0,0)(38,-15){-2}{3} \Text(5,-10)[lt]{\footnotesize$\;\gamma$}   
\SetWidth{1} \BCirc(-40,10){25}    
\SetWidth{1} \BCirc(45,10){25}    
\Text(-50,10)[lc]{\footnotesize$(\pi\pi)_I\;$} 
\Text(32,10)[lc]{\footnotesize$\pi^+\pi^-\gamma\;$} 
\end{picture} 
   \hspace*{\fill} 
\begin{picture}(80,50)(-40,-15)    
\Line(-20,20)(18,20) \Text(20,20)[lc]{\footnotesize$\pi^+\;$}\Line(32,20)(70,20)  
\Line(-20,0)(18,0) \Text(20,0)[lc]{\footnotesize$\pi^-\;$} \Line(32,0)(70,0)  
\Photon(0,0)(20,-20){-2}{3} \Text(20,-20)[lt]{\footnotesize$\;\gamma$}   
\Photon(30,-20)(50,0){-2}{3} 
\SetWidth{1} \BCirc(-40,10){25}    
\SetWidth{1} \BCirc(90,10){25}    
\Text(-50,10)[lc]{\footnotesize$(\pi\pi)_0\;$} 
\Text(80,10)[lc]{\footnotesize$(\pi\pi)_2\;$} 
\end{picture} 
   \hspace*{\fill} 
\caption{\label{figone}%
A schematic illustration of how electromagnetism can generate 
channel-coupling effects. We let a round circle represent
a particular final-state channel.  
The $(\pi\pi)_I \leftrightarrow \pi^+\pi^-\gamma$ channel
coupling starts in ${\cal O}(e)$, whereas the 
$(\pi\pi)_0\leftrightarrow (\pi\pi)_2$ channel coupling, mediated
by the $\pi^+\pi^-\gamma$ channel, starts in 
${\cal O}(e^2)$. A 
$(\pi\pi)_0\leftrightarrow (\pi\pi)_2$ channel coupling 
can also be mediated by photon exchange, so that the
presence of an intermediate $\pi^+\pi^-\gamma$ state is 
not necessary. 
}  
\end{figure}               

\begin{itemize}
\item[1)] 
We assume that the $\pi^+ \pi^- \gamma$
final state is the {\em only} inelastic channel which couples 
to the two--pion channel of given isospin, thus generating 
transitions of the type 
$(\pi \pi)_0 \leftrightarrow  (\pi \pi)_2$ as illustrated
in Fig.~1. 
As we have noted, 
the inelasticities generated by the
opening of the four--pion threshold can be safely neglected. 
Considering the empirical branching ratios of $K_S$ to electromagnetic
final states, we note that the branching ratio for 
$K_S \to \pi^+\pi^- \gamma$, presuming photon momenta in
excess of 50 MeV/c\footnote{The cut on the photon momentum is required; 
the $\pi^+\pi^- \gamma$ final state generated by 
bremsstrahlung from a charged pion is infrared divergent.}, 
is roughly $2 \cdot 10^{-3}$, whereas the next largest
measured branching ratio, $K_S \to \gamma \gamma$, is roughly
a factor of 1000 smaller~\cite{pdg}. 
$K_S \to 3 \pi$ decay is also possible, but
is unimportant in this context, as 
a $3\pi \to 2\pi$ transition with $J=0$ 
violates parity and cannot contribute in ${\cal O}(G_F)$ 
to the unitarity relations of interest. 
Consequently, the $\pi^+\pi^- \gamma$ final state is the
only inelastic channel of interest, and 
the appropriate extension 
of Eq.~(\ref{master3}) 
is a 4$\times$4 matrix.
\item[2)]We  work in leading order in the Fermi constant, ${\cal
O}(G_F)$, and focus on the unitarity constraints in which
the kaon decay amplitudes, $A_I$ for $K\to (\pi\pi)_I$ decay and
$A_\gamma$ for $K\to \pi^+\pi^- \gamma$ decay, 
 appear. Thus we may choose  
the 3$\times$3  submatrix without kaons, that is, 
that of the coupled $(\pi\pi)_0$,  $(\pi\pi)_2$, and $\pi^+ \pi^- \gamma$
system, which we term
the pion--photon system, to be both unitary 
and T--reversal--invariant. 
The $3\times3$ matrix contains eighteen parameters; 
nine parameters are constrained by unitarity, and three more are
constrained by T 
invariance, so that the resulting matrix
has {\it six} non--trivial parameters.
Scattering in these channels is driven by strong and electromagnetic effects.
\item[3)] Recalling the form of the bar phase shifts in 
NN scattering~\cite{preston}, 
we write the 3$\times$3 S--matrix of the pion--photon system in 
an analogous form,
${\cal A} \cdot {\cal B} \cdot {\cal A}$, where ${\cal A}$ 
is a diagonal matrix parametrized in terms
of the phase shifts of $\pi\pi$ and $\pi^+\pi^-\gamma$ scattering,
namely $\bar\delta_I$ and $\bar\delta_\gamma$, so that
\beqa\label{A}
{\cal A} =
\left( \matrix {  e^{i\bdg} & 0 &  0 \nonumber \\
 0 & e^{i\bdz}  & 0  \nonumber \\
 0 & 0  & e^{i\bdt}   
     \nonumber\\} \!\!\!\!\!\!\!\!\!\!\!\!\!\!\!\!\!\!  \right) 
 \,\, ,
\eeqa
cf. Eq.~(\ref{Sbar}), and
${\cal B}$ is a unitary, T-invariant 3$\times$3 matrix 
containing 
three parameters. 
Our form of ${\cal B}$  is inspired by the form of the 
Kobayashi-Maskawa 
parametrization of the 
Cabibbo-Kobayashi-Maskawa matrix~\cite{KM}. 
The latter, however, contains 4 parameters, but one can readily 
define one of the parameters in
terms of the others to yield a T--invariant matrix. More precisely,
we introduce two angles $\Theta_{1,2}$ and one phase $\delta$. This
assignment will be discussed below. The matrix ${\cal B}$ is chosen to be:
\beqa\label{B}
{\cal B} =
\left( \matrix {  c_1 & i s_1 \, c_2  & i s_1 \, s_2 
     \nonumber \\
                 i s_1 \, c_2 &  c_1 \, c_2^2 + s_2^2 e^{i\delta} 
                  & c_2 \, s_2 (c_1 - e^{i\delta} ) 
     \nonumber \\
                 i s_1 \, s_2 & c_2 \, s_2 (c_1 -  e^{i\delta})
                 & c_1\, s_2^2 +  c_2^2 e^{i\delta} 
     \nonumber\\} \!\!\!\!\!\!\!\!\!\!\!\!\!\!\!\!\!\!  \right) 
 \,\, ,
\eeqa
where we adopt the conventional abbreviations $s_i \equiv \sin \Theta_i$ and 
$c_i \equiv \cos \Theta_i$.
\item[4)] The form of $B$ in Eq.~(\ref{B}) is compatible with the hierachy
of channel couplings. The transitions 
$ (\pi^+ \pi^- \gamma) \leftrightarrow 
(\pi \pi)_I$ start at ${\cal O}(e)$, whereas the couplings between the 
two--pion final states  $(\pi \pi)_0 \leftrightarrow (\pi \pi)_2$ are of
${\cal O}(e^2)$. 
It is seemly, then, that $\Theta_1$ appears as $s_1$ in the
$ (\pi^+ \pi^- \gamma) \leftrightarrow (\pi \pi)_I$ elements
 and as $c_1$ in the $(\pi \pi)_0 \leftrightarrow (\pi \pi)_2$ elements. 
Furthermore, at ${\cal O}((m_u-m_d)^2)$, there are no
transitions between $(\pi\pi)_I$ and $\pi^+\pi^- \gamma$, whereas there
are transitions between 
$(\pi\pi)_I$ and  $(\pi\pi)_{I'}$. 
Thus the introduction
of the phase 
$\delta$ is convenient as this {\it can} describe the presence of 
$m_d \ne m_u$ effects as distinct from the 
electromagnetic isospin--violating effects characterized by 
$\Theta_1$. Of course $\delta$ can contain electromagnetic 
contributions as well. 
Note that 
taking $\Theta_1 =0$, $\delta =0$ sets {\em all} the 
channel couplings to zero. Moreover, including the parameter $\delta$ as
a phase means that in the limit in which only 
${\cal O}(\delta,\Theta_1)$ terms are kept, all the off-diagonal
terms are imaginary. 
The remaining parameter, $\Theta_2$, can
be thought of as characterizing the difference of the 
inelasticity parameters in the $I=0$ and $I=2$ channels  due to
the presence of the third channel.  
\item[5)] 
We work with IR-finite
amplitudes throughout. 
For a detailed discussion of the extraction of the IR-finite
parts from the full amplitudes, we refer the reader to 
Ref.~\cite{CDGIII}. Here,
it suffices to say that the potentially troublesome contributions
can be factored and are thus not relevant to our discussion.
\end{itemize}
We can now
give the generalized 4$\times$4 matrix, where the
$K^0$, $\pi^+\pi^-\gamma$, $(\pi\pi)_0$, and  $(\pi\pi)_2$ channels are
associated with rows/columns 1,2,3, and 4, respectively. This assignment 
is prompted by the foregoing remarks. We thus have
\beqa\label{master}
{\cal S} = 
\left( \matrix { 1 & iA^*_\gamma \, e^{i\tdg} 
     & iA^*_0 \, e^{i\tdz} &  iA^*_2 \, e^{i\tdt} 
     \nonumber \\ 
                 iA_\gamma \, e^{i\tdg} & c_1 \, e^{2i\bdg}
     &  is_1 \, c_2 \, e^{i(\bdz+\bdg)} & is_1 \, s_2 \, e^{i(\bdg+\bdt)}
     \nonumber \\
                 iA_0 \, e^{i\tdz} & i s_1 \, c_2 \, e^{i(\bdz+\bdg)}
     & ( c_1 \, c_2^2 + s_2^2 e^{i\delta}) \, e^{2i\bdz} 
     & c_2 \, s_2 (c_1 - e^{i\delta} ) \, e^{i(\bdz+\bdt)}
     \nonumber \\
                 iA_2 \, e^{i\tdt} & i s_1 \, s_2 \, e^{i(\bdg+\bdt)}
     & c_2 \, s_2 (c_1 -  e^{i\delta}) \, e^{i(\bdz+\bdt)} 
     & (c_1\, s_2^2 + c_2^2 e^{i\delta} ) \, e^{2i\bdt}
     \nonumber\\} \!\!\!\!\!\!\!\!\!\!\!\!\!\!\!\!\!\!  \right) 
 \,\, .
\eeqa
Based on this matrix, we can now derive the consequences of the 
extension of Watson's theorem including electromagnetism and
strong-interaction isospin violation. 
To the best of our knowledge, this form 
of the unitarity constraints has not appeared previously in the literature.

\medskip\noindent
Before working out the unitarity constraints realized from 
this 4$\times$4 matrix, 
we briefly discuss the relation of its  2$\times$2 submatrix for 
$\pi\pi$ scattering to existing parametrizations in the literature. 
Were the inelastic channel not present, three parameters would 
suffice in characterizing it. 
We have five parameters, whereas 
the following parametrization for the $\pi\pi$
transition matrix in the isospin basis is proposed in Ref.~\cite{CDGIII}:
\beqa\label{MCDG}
\sqrt{1- 4M_\pi^2 / M_K^2 }\, {\overline T}_{\rm iso} 
= \left( \matrix { \frac{1}{2i}\left( \eta_0 e^{2i \delta_0} - 1 \right) & 
   a \, e^{i(\delta_0+\delta_2 + \Delta)} \nonumber \\
   a \, e^{i(\delta_0+\delta_2 + \Delta)} 
 & \frac{1}{2i}\left( \eta_2 e^{2i \delta_2} - 1 \right)
 }  
                           \!\!\!\!\!\!\!\!\!\!\!\!\!\!\!\!\!\!\! \right) 
 \,\, .
\eeqa
Only T
invariance is imposed, and the ``bar'' denotes 
the IR finite amplitudes. The parameter $a$ controls the isospin
mixing, and the opening of possible new channels, yielding 
a violation of unitarity in the $(\pi\pi)_I$ sector, 
is parametrized in terms of the inelasticity 
parameters $\eta_0$ and $\eta_2$. The assumption
of having only one additional open channel ($\pi^+\pi^-\gamma$) leads 
to a correlation
between $\eta_0$ and $\eta_2$, as seen in Eq.~(\ref{master}). 
We should also note that
in the limit $\Theta_1 = 0$, our 2$\times$2 submatrix for the $\pi\pi$ system
is characterized by four parameters, so that 
one of the parameters is redundant, as the resulting 
submatrix is both unitary and T--reversal--invariant. The resulting
relations between the parameters of this 2$\times$2 matrix 
and those of Eq.~(\ref{Sbar}) can be determined, but are not 
transparent. 
\medskip

\noindent {\bf 4.} 
We  proceed to derive the explicit form of the {\em unitarity constraints}
from Eq.~(\ref{master}). Specifically, 
$({\cal S}^\dagger {\cal S})_{21} = ({\cal S}^\dagger {\cal S})_{31}
=  ({\cal S}^\dagger {\cal S})_{41} = 0$ yields
\beqa\label{Eqstar1}
A_\gamma &=& A_\gamma \, c_1 \, e^{2i (\Dg)} - A_0 \, i s_1 \, c_2 \,
e^{i(\Dz+\Dg)} - A_2 \, i s_1\, s_2 e^{i(\Dt+\Dg)}~,\\
A_0 &=& - A_\gamma i s_1\, c_2\, e^{i(\Dz+\Dg)} 
+ A_0 \, (c_1 \, c_2^2 +s_2^2 \, e^{-i\delta}) \,
e^{2i (\Dz)} + A_2\, c_2\, s_2 \, (c_1 -  e^{-i\delta}) \,
 e^{i(\Dz + \Dt)}~,\nonumber\label{Eqstar2} \\ && \\
A_2 &=& - A_\gamma i s_1\, s_2\, e^{i(\Dt + \Dg)} 
+ A_0 \,  c_2\, s_2 \, (c_1 -  e^{-i\delta}) \,  e^{i(\Dz + \Dt)}
+A_2 \, (c_1 \, s_2^2 +c_2^2 \, e^{-i\delta}) \, e^{2i (\Dt)}~.
\nonumber \\ &&\label{Eqstar3} 
\eeqa
For $\Theta_1 = \delta = 0$ we recover $\tilde{\delta}_0 = 
\bar{\delta}_0$, $\tilde{\delta}_2 = \bar{\delta}_2$, and
$\tilde{\delta}_\gamma = \bar{\delta}_\gamma$ as we
would expect. Recalling Eq.~(\ref{Deldef}) and defining
$\Delta_\gamma=\tdg - \bdg$, we proceed 
by eliminating $A_\gamma \exp(i\Deg)$ from 
Eqs.~(\ref{Eqstar2},\ref{Eqstar3}) to recover
\beq\label{cons}
  A_0 \, s_2 \, \left( 2i \, \sin\Dez + (e^{-i\delta}-1)\, 
e^{i\Dez} \right)
- A_2 \, c_2 \, \left( 2i \, \sin\Det + (e^{-i\delta}-1)\, 
e^{i\Det} \right)
= 0~.
\eeq
Remarkably these formulae do not depend on 
$\Theta_1$, and we can factor $e^{-i \delta/2}$ 
to obtain
\beq\label{beaut}
\frac{A_2}{A_0}= \tan \Theta_2 \, \left(
\frac{\sin(\Dez - \delta/2)}{\sin(\Det - \delta/2)}
\right)\;.
\eeq
We find, as discussed previously, that $\Det \gg
\Dez$ since the $I=2$ phase shift is enhanced by a factor of 
$A_0/A_2\sim 22$~\cite{GV,CDGIII}. 
Moreover, no terms in $\Theta_1$ appear, so that the explicit
coupling to the $\pi^+\pi^-\gamma$ channel is irrelevant to 
$\Delta_I$. Interestingly, $\Theta_1$ does not enter  
either Eq.~(\ref{cons}) or Eq.~(\ref{beaut}). 
This was also observed in Ref.~\cite{CDGIII}, but arose
from the results of a numerical analysis, whereas here it emerges
as a consequence of unitarity. Let us make one more comment
about Eq.~(\ref{beaut}) before proceeding. The right-hand side
is explicitly real, whereas the left-hand side is not, as the
amplitudes $A_I$ carry weak phase information. Requiring the
imaginary part of the left-hand side 
of Eq.~(\ref{beaut}) to be zero yields the constraint
$\rm{Im}\,A_2/\rm{Re}\,A_2 - \rm{Im}\,A_0/\rm{Re}\,A_0=0$, 
apparently suggesting that the CP-violating parameter
Re$\,(\epsilon'/\epsilon)$ is zero as a consequence of unitarity. 
However, this is not the case: the amplitudes in $A_I$ contain
isospin violation as well, and  
Eq.~(\ref{beaut}) becomes
indefinite in the isospin-perfect limit.  
If we proceed to examine the relationship between $A_I$ and 
the amplitudes appearing in the general parametrization of Ref.~\cite{GV} of
the isospin-breaking effects in the $K\to\pi\pi$ amplitudes, 
we find that ${\rm Re}\,A_I$ and ${\rm Im}\,A_I$ 
are distinguished by an additional, isospin-violating, CP-conserving 
function. Physically this implies that our parametrization of the
${\cal S}_{i1}$ and ${\cal S}_{1i}$ matrix elements is not sufficiently
general, that in the presence of isospin violation, there is an
additional, path-dependent CP-conserving piece. For our purposes
we can neglect this additional contribution, though it could
well impact the Standard Model prediction of Re$\,(\epsilon'/\epsilon)$, 
and we thus proceed to 
neglect weak phases throughout. 
Thus we interpret $A_\gamma$ and 
$A_I$ as ${\rm Re}\,A_\gamma$ and as ${\rm Re}\,A_I$, respectively. 

\medskip\noindent
We can also eliminate $A_2$ from Eqs.~(\ref{Eqstar1},\ref{Eqstar2})
to yield
\beq\label{gamma0}
 A_\gamma \, c_2 \, \left[
 \cos(\Deg + \delta/2) - c_1 \cos(\Deg - \delta/2)
\right]
= A_0\,s_1 \, \sin( \Dez - \delta/2 )\;,
\eeq
and, similarly, $A_0$  from Eqs.~(\ref{Eqstar1},\ref{Eqstar3}),
\beq\label{gamma2}
 A_\gamma \, s_2 \, \left[
 \cos(\Deg + \delta/2) - c_1 \cos(\Deg - \delta/2)
\right]
= A_2\,s_1 \, \sin( \Det - \delta/2 )\;.
\eeq
It is worth noting that in Eqs.~(\ref{gamma0},\ref{gamma2}) 
the parameter $\Theta_1$ does explicitly appear, controlling
the relation between $A_\gamma$ and $A_I$. Combining these
latter two equations yields Eq.~(\ref{beaut}), as it should.
Since the $A_I$ and $A_\gamma$ are complex, our 
remarks concerning the most general parametrization of 
$A_I$ apply to the $A_\gamma$ amplitude as well. 

\medskip\noindent
We also point out  that $\Delta_\gamma$ itself is only
non-zero in ${\cal O}(e^3)$, $\tilde \delta_\gamma$ being
given by the $I=1,L=1$ $\pi\pi$ phase shift, 
so that additional expressions are possible. We refrain from
reporting these. 
Nevertheless, we have extended 
Watson's theorem 
to include the presence
of electromagnetism, for the 
parameters of $(\pi\pi)_I$ and $\pi^+\pi^-\gamma$
scattering suffice to relate the electromagnetically generated phases in
$K \to \pi\pi$. 
 
\medskip

\noindent {\bf 5.} 
In this letter, we have considered the unitarity constraints on the
strong and electromagnetically induced phases in $K\to \pi\pi$ decays.
Assuming that the $(\pi^+\pi^-\gamma)$ final state is the only
inelastic channel, and working in ${\cal O}(G_F)$, we have derived the
constraints between the decay amplitudes $A_I$ for $K\to
(\pi\pi)_I$ decay (for isospin $I=0,2$) 
and $A_\gamma$ for $K \to \pi^+\pi^-\gamma$. The
corresponding S--matrix is characterized by three mixing parameters; 
we choose two angles $\Theta_{1,2}$ and one phase $\delta$. The
angle $\Theta_1$ is chiefly responsible for the channel couplings
$(\pi^+\pi^-\gamma) \leftrightarrow (\pi\pi)_I$, whereas $\delta$
describes the strong isospin violation effects due to the light quark
mass difference, as well as any ``direct'' electromagnetic
coupling between the $(\pi\pi)_0 \leftrightarrow (\pi\pi)_2$ states.
{}From the general $4\times 4$ S-matrix for $K^0$
decays into these three final states, cf. Eq.(\ref{master}), one
can derive a set of unitarity constraints. Most remarkably, it can
be shown that the explicit coupling of the
$(\pi^+\pi^-\gamma)$--channel is irrelevant to the difference between
the T-conserving 
$\pi\pi$ phases measured in $\pi\pi$ scattering and
extracted from $K\to \pi\pi$ 
(and $K\to\pi^+\pi^-\gamma$\footnote{Recall that we work with IR 
finite amplitudes throughout.}) 
decays. 
We have also argued that in the
presence of isospin violation, the 
parametrization for the complex-valued decay amplitudes $A_I$ ought 
be modified;
we illuminate the 
sources of isospin breaking which give rise to the
general parametrization of Ref.~\cite{GV}.
As a next step, it will be interesting to work out the 
numerical consequences of these constraints on the determination
of the $|\Delta I|=5/2$ amplitude in 
 $K \to \pi\pi$ 
decay, as well as the
impact on the Standard Model prediction for the CP-violating 
parameter Re$\,(\epsilon'/\epsilon)$.

\vskip 1cm

\noindent{\large {\bf Acknowledgements}}

\smallskip\noindent
S.G. and U.-G.M. thank the Institute of Nuclear Theory at the University of
Washington for its hospitality during the completion of this work.  
The work of S.G. and G.V. is supported in
part by the U.S. Department of Energy under contract numbers 
DE-FG02-96ER40989 and DE-FG02-01ER41155, respectively.



\vskip 1cm

\begin{thebibliography}{99}

\frenchspacing
\bibitem{GMI}J. Gasser and Ulf-G. Mei{\ss}ner, Phys. Lett. B 258 (1991) 219.\vs
\bibitem{Bern} B. Ananthanarayan et al., [hep-ph/0005297].\vs
\bibitem{EdR}E. de Rafael, Nucl. Phys. B (Proc. Suppl.) 7 (1989) 1.\vs
\bibitem{CDGI} V. Cirigliano, J.F. Donoghue, and E. Golowich,
  Phys. Rev. D 61 (2000) 093001 [hep-ph/9907341].\vs
\bibitem{CDGII} V. Cirigliano, J.F. Donoghue, and E. Golowich,
  Phys. Rev. D 61 (2000) 093002 [hep-ph/9909473].\vs
\bibitem{CDGIII} V. Cirigliano, J.F. Donoghue, and E. Golowich,
  Eur. Phys. J. C 18 (2000) 83 [hep-ph/0008290].\vs
\bibitem{Gardner:1999wb}
S.~Gardner and G.~Valencia,
Phys.\ Lett.\ B 466 (1999) 355.
[hep-ph/9909202].\vs
\bibitem{GV} S. Gardner and G. Valencia, Phys. Rev. D 62 (2000)
  094024 [hep-ph/0006240].\vs
\bibitem{WienI}G. Ecker et al., Nucl. Phys. B 591 (2000) 419 [hep-ph/0006172].\vs
\bibitem{WienII}G. Ecker et al., Phys. Lett. B 477 (2000) 88
  [hep-ph/9912264] . \vs
\bibitem{aron}A.M. Bernstein, Phys. Lett. B 442 (1998) 20.\vs
\bibitem{MMS}B.R. Martin, D. Morgan, and G. Shaw, {\em Pion-Pion
  Interactions in Particle Physics}, Academic Press, London, 1977.\vs
\bibitem{JLP}J.L. Petersen, CERN Yellow Report 77-04, 1977.\vs
\bibitem{GMII} J. Gasser and Ulf-G. Mei{\ss}ner, Nucl. Phys. B 357
  (1991) 91.\vs
\bibitem{preston} 
H.P. Stapp, T.J. Ypsilantis, and N. Metropolis, Phys. Rev. 105 (1957) 302.\vs
\bibitem{GL} J. Gasser and H. Leutwyler, Ann. Phys.  158  (1984) 142.\vs
\bibitem{pdg} D.E. Groom {\it et al.} (Particle Data Group),
Eur. Phys. J. C 15 (2000) 1. \vs
\bibitem{KM}M. Kobayashi and T. Maskawa, Prog. Theor. Phys. 49 (1973) 652.\vs

\end{thebibliography}
\end{document}